# The Universe, the Cold War, and Dialectical Materialism

HELGE KRAGH*

**Abstract**. Ideological considerations have always influenced science, but rarely as directly and massively as in the Soviet Union during the early Cold War period. Cosmology was among the sciences that became heavily politicized and forced to conform to the doctrines of Marxism-Leninism. This field of science developed entirely differently in the Communist countries than in the West, in large measure because of political pressure. Certain cosmological models, in particular of the big bang type, were declared pseudo-scientific and idealistic because they implied a cosmic creation, a concept which was taken to be religious. The result of the ideological pressure was not an independent Soviet cosmology, but that astronomers and physicists abandoned cosmological research in the Western sense. Only in the 1960s did this situation change, and cosmology in the Soviet Union began to flourish. The paper examines the relationship between science and political ideology in the case of the Soviet Union from about 1947 to 1963, and it also relates this case to the later one in the People's Republic of China.

## 1. *Introduction: Stalinism and the Sciences*

The Cold War was not only a confrontation between two antagonistic political systems that involved military, political and economic actions; it was also a confrontation between two world views in which science and philosophy, directly and indirectly, were parts of the political agenda. In some cases scientific theories became politicized, that is, associated with political and ideological views that made them either attractive or unattractive. In the latter case they might be judged politically incorrect to such an extent that they effectively became suppressed as *theoria non grata*. In the Western propaganda, scientific views were occasionally associated with Marxist values, such as materialism and atheism, which made it easier to discredit them and question their scientific legitimacy.

---

* Centre for Science Studies, Institute of Physics and Astronomy, Aarhus University, Building 1520, 8000 Aarhus, Denmark. E-mail: helge.kragh@ivs.au.dk. Manuscript submitted to the journal *Centaurus*.

However, it was only within the authoritarian system of the Soviet Union and its allied nations that scientific theories were directly suppressed for political reasons, such as happened most flagrantly in the era of Stalinism from about 1946 to 1953 (Graham, 1972; Pollock, 2006). According to the Central Committee of the Communist Party, it was the duty of every Soviet citizen to 'defend the purity of Marxist-Leninist doctrines in all domains of culture and science' (Prokofieva, 1950, p. 12). Moreover, the Committee stressed that science is not cosmopolitan, but divided along the line of the world-wide class struggle, with a materialistic Soviet science fighting the idealistic pseudo-science of the capitalist world.

The question of the relationship between Communist ideology and scientific thought in the early phase of the Cold War is complex, for other reasons because the severely repressive political system did *not* cause a general decline in Soviet science. On the contrary, during the same period science in the Soviet Union made remarkable advances, a phenomenon that Alexei Kojevnikov (2004, p. xii) has called the 'main paradox of Soviet science'. The infamous Lysenko affair and the crusade against Western genetics did have damaging consequences for Soviet biology and agricultural science, but Lysenkoism was hardly the symbol of the ideology-science relationship that it has often been made (Krementsov, 1997; Kojevnikov, 2004, pp. 186-214). At any rate, it was on a much bigger scale than the suppression of cosmological thought here considered. There are other cases more comparable to what happened in cosmology, such as the relatively little known controversy in structural chemistry focusing on the resonance theory of aromatic compounds. Briefly, in this case the quantum theory of resonance was considered politically incorrect because it did not describe molecules as real structures in accordance with the sanctioned view of materialism (Graham, 1964; Pechenkin, 1995).[1]

Andrei Zhdanov, a member of the Politburo and Stalin's chief ideologue, was the driving force behind the political alignment of culture and science, including the purge of incorrect views from Soviet science. On 24 June 1947 he delivered a speech in which he condemned trends in philosophy and science that he deemed contrary to the values of Marxism-Leninism. Astronomy and cosmology were among the sciences that needed to be cleansed of bourgeois heresies. Referring to 'the reactionary scientists Lemaître, Milne and others', Zhdanov accused Western cosmology of being covertly religious. It used the observed redshifts of the nebulae 'to strengthen religious views on the structure of the universe', he said. Moreover, 'Falsifiers of science want to revive the fairy tale of the origin of the world from nothing. … Another failure of the "theory" in question consists in the fact that it brings us to the idealistic attitude of assuming the world to be finite'.[2] Zhdanov's talk marked the beginning of a decade in which cosmology in the sense cultivated by Western physicists and astronomers almost disappeared from Soviet science, in large measure because it was seen as politically incorrect. Compared with the situation in other parts of the physical sciences, such as quantum mechanics and relativity theory, this is a case that has only attracted limited attention from historians of science (but see Graham, 1972, pp. 139-194; Haley, 1980; Kragh, 1996, pp. 259-268).

## 2. *Shadows of the Past: Engels to Lenin*

What in the early 1950s emerged as the doctrines of Communist cosmology, as defined by orthodox party philosophers, can be summarized in five points (Haley, 1980, pp. 139-149; Mikulak, 1958):

(i) The universe is infinite in both space and its content of matter.

(ii) The universe is eternal: there never was a beginning and there never will be an end.

(iii) Only matter and its manifestations in the forms of motion and energy have any real existence in the universe.

(iv) The truth of cosmological theories should be judged by their correspondence with the laws of dialectical-materialist philosophy.

(v) The galactic redshifts do not indicate that cosmic space is in a state of expansion, but can be explained by other mechanisms.

Remarkably, the first four of the doctrines have their roots in the nineteenth century and are essentially repetitions of what Friedrich Engels, Marx's close collaborator, argued in his works on the dialectics of nature. (He was unaware of the expansion of the universe, which was only discovered in the late 1920s.) To understand the position of the Stalinist party philosophers it is necessary to briefly recall the ideological discussion in the late nineteenth century concerning thermodynamics and cosmology.

The idea of a cosmic 'heat death' caused by the continual increase of entropy in the universe, which is one version of the second law of thermodynamics, was intensely discussed in the second half of the nineteenth century by scientists and non-scientists alike (Neswald, 2006; Kragh, 2008). Many philosophers and social critics, including the large majority of socialist thinkers, found it unbearable that life and activity in the universe should one day cease to exist. They found it equally unacceptable that it apparently followed from the second law that the universe had a finite age; because, if it were infinitely old, the entropy would be maximally high, contrary to observation. Then, from a universe of finite age there was but a short step to one which was created supernaturally. What matters in the present context is that socialists generally rejected both the heat death and the 'entropic creation argument'. One way of escaping these unpalatable consequences was to postulate an infinitely large universe to which the laws of thermodynamics supposedly did not apply.

Engels was much worried of the prospects of a running-down universe, which he argued against in his *Dialektik der Natur* and other works on the dialectical natural philosophy. For example, in a letter to Karl Marx of 21 March 1869 he claimed that the heat death scenario was not only scientifically nonsense but also ideologically dangerous: 'Since, according to this theory, in the existing world, more heat must always be converted into other energy than can be obtained by converting other energy into heat, so the original *hot state*, out of which things have cooled, is obviously inexplicable, *even contradictory*, and thus presumes a God' (Kragh, 2008, p. 135). To the mind of the militantly atheistic Engels, cosmic irreversibility was incompatible with dialectical materialism, whereas it legitimated miracles and divine creation. The universe must necessarily be a *perpetuum mobile*, and for this reason infinite in matter and space. This view, generally accepted by early thinkers of a socialist and positivist inclination, was adopted by Lenin and canonized in his philosophical treatise *Materialismus und Empiriokritizismus* from 1908. Considered to be an integral part of the doctrines of dialectical materialism, it was incorporated in the official philosophy of nature that came to dominate thinking in the Soviet Union and other Communist countries.

Well before the Cold War, party philosophers in the Soviet Union had taken Engels' cosmic thoughts to their hearts and turned them into doctrines of the Communist world view. To repeat, according to this world view the universe was infinite and eternal, self-regulating and in eternal flux. The dogma even became enshrined in the officially approved definitions of cosmology. One such definition, dating from the early 1950s, reads (Hayes, 1980, p. 151):

> Cosmology is the study of an infinite universe as a coherent, single whole and of the whole region embraced by observation as a part of the universe. This study has … the status of an independent branch of astronomy, closely associated with physics. In its generalization, cosmology is essentially governed by philosophy and cannot be scientific without a philosophical base containing a correct theory of knowledge and revealing general laws of matter and of its motion.

Soviet ideologues considered the very application of physical theories to the universe as a whole to be suspect and un-Marxist as long as these theories were not 'governed by philosophy' – meaning dialectical materialism. They found it unscientific as well as ideologically unacceptable to extrapolate local laws of physics, such as relativity theory and thermodynamics, to the universe at large.

The cosmological consequences of the second law of thermodynamics were discussed on both sides of the Iron Curtain, but in different ways. In the 1950s the heat death did not occupy an important position among Western cosmologists, who chose to focus on properties of the universe that could be determined observationally, such as its space curvature and expansion rate. The heat death was considered too hypothetical to be of use in discriminating between cosmological models. On the other hand, Soviet scientists and philosophers took the subject very seriously, in part motivated by the political consensus that the universe could not possibly end in an equilibrium state. The answer was given in advance, dictated by the philosophical system. They consequently adopted various strategies to refute the idealistic heat death hypothesis (Graham, 1972, p. 500). The strategies were basically the same as used in the nineteenth century, the most popular being to deny that the law of entropy increase applied to the entire universe or to suggest the existence of processes counteracting the growth in entropy. For example, in 1950 the physicist J. R. Plotkin argued that 'a state of equilibrium for the whole universe not only is impossible, but does not make any sense at all. … Attempts at applying to the whole universe the conclusion of the second law of thermodynamics have no scientific foundation'.[3]

## 3. *Religion and Cosmological Theories*

The cosmological scene in the early 1950s was confusing, with no consensus model of the universe and no agreement about the proper methods of cosmology as a science (North, 1965; Kragh, 1996). In Kuhnian terms, cosmology was lacking a paradigm and therefore still in a prescientific stage. The majority of physicists and astronomers on both sides of the Iron Curtain agreed that the universe expands and that the expansion was best explained by Einstein's cosmological field equations, although in the Soviet Union the term 'universe' was typically understood in a different sense than in the capitalist countries. To mainstream cosmologists in the West, a cosmological model corresponded to a solution of Einstein's equations, and the problem was to find by means of theory and observation the model describing the one and only real universe. Many relativistic cosmologists were in favour of the ever-expanding Lemaître-Eddington model, which had no sudden beginning in time; others found a finite-age universe of the explosive (big bang) type to be an attractive possibility. The idea of a big bang universe was first proposed by the Belgian cosmologist and Catholic priest Georges Lemaître in 1931, without attracting much attention. In the period 1946-1953 it was turned into a physical model of the early universe by the Russian-American nuclear physicist George Gamow and his collaborators Ralph Alpher and Robert Herman, but also this model failed to win acceptance.

Then there were cosmological theories that were not based on the general theory of relativity and did not assume a universe of finite age. The most important of these alternatives was the steady state theory introduced by Fred Hoyle, Hermann Bondi and Thomas Gold in 1948. According to this theory, the universe expanded at an ever increasing rate and yet it had always looked the same and would continue to do so; there was neither a beginning nor an end of time. To keep the universe in a steady state it was necessary to postulate that matter was continually created throughout the universe, a feature that contributed to make the theory controversial. From 1948 to about 1965, when the steady state theory was largely abandoned, it was involved in an epic controversy with the rival class of relativistic evolution theories (Kragh, 1996). All this took place in the Western world, mostly in England and the United States, whereas none of the two classes of cosmological theories found approval in the Soviet bloc.

Religion was a most important element in the war over the souls that was an integral part of the Cold War. The authorized Soviet version of dialectical materialism was radically opposed to religion and obliged to fight it in whatever of its manifestations, including its associations to science. Astronomy had served as a vehicle for Soviet anti-religious propaganda also before the war, when the Catholic Church was under ideological attack. The propaganda resulted in a brief but interesting controversy between Russian Communist astronomers and Otto Struve, the eminent Russian-American astronomer and director of the Yerkes Observatory (Struve, 1935; Bronshten and McCutcheon, 1995).

Communist party philosophers saw an unholy alliance between the Christian doctrine of genesis and the finite-age models proposed by cosmologists such as Gamow and Lemaître. In the case of Gamow the suspicion was unfounded, as Gamow was not a Christian but either an agnostic or an atheist. It was equally unfounded in the case of Lemaître, who was careful to distinguish between the 'beginning' and the 'creation' of the world. According to Lemaître (1958, p. 7), his version of the big bang model 'remains entirely outside any metaphysical or religious question [and] leaves the materialist free to deny any transcendental Being'. Nevertheless, it was commonly claimed that Lemaître's explosive universe was apologetically motivated. The accusation was routinely made in Soviet comments, but it or similar claims can be found also in non-Marxist Western scientists and philosophers.[4] While unjustified in the case of Lemaître, apologetic uses of big bang cosmology were sometimes made in the period, confirming atheist and socialist critics in their belief that the big bang theory was a religious view masquerading as science.

The most remarkable and publicized example of such misuse was an official, so-called encyclical address that the pope, Pius XII, gave in Rome on 22 November 1951. In this much-discussed address the pope effectively argued that the new and still hypothetical big bang theory served as scientific legitimation for what the faithful had always known, that the universe was created by God. Whereas the big bang theory was at the time a minority view without convincing evidence, the pope's address gave the false impression that it was the authoritative theory of the universe supported by most scientists. Present-day science, the pope said, 'have succeeded in bearing witness to the august instant of the primordial *Fiat Lux*, when, along with matter, there burst forth from nothing a sea of light and radiation, and the elements split and churned and formed into millions of galaxies' (Pius XII, 1951; McLaughlin, 1957, pp. 137-147). The modern physical theory of the universe, he concluded, has 'confirmed the contingency of the universe and also the well-founded deduction as to the epoch when the world came forth from the hands of the Creator. Hence, creation took place. We say: therefore, there is a Creator. Therefore, God exists!'.

The following year, 1952, the General Assembly of the International Astronomical Union (IAU) took place in Rome with 429 delegates from 35 countries. The meeting illustrates the difficulties of maintaining internationalism and scientific cooperation under the conditions of the Cold War, but also the wish of the scientists to do so. IAU was the only international scientific union to which the Soviet Union belonged at the time and thus of particular importance to the scientists. The Rome meeting had originally been planned to take place in Leningrad, but had been cancelled by the IAU Executive Committee for political reasons, primarily the Korean War and the heightened tension between East and West. Adding to the decision was the persistent Communist propaganda against 'bourgeois astronomy' and doubts about the freedom of Soviet astronomers. Having returned from a meeting in the Polish Academy of Sciences, in 1948 the British astronomer and former President of IAU, Harold Spencer Jones, wrote to the new General Secretary of IAU, the Dane Bengt Strömgren: 'There is a very widespread concern about the present trends and the threats to scientific and intellectual freedom, which no doubt derive from orders from Moscow. One of the Russian delegates gave an address to the Academy on the work of Lysenko and emphasized that genetical thought must develop along Marxist lines. I am aware also that cosmogony is being subjected to party pressure' (Blaauw, 1994, p. 165).

The Soviet delegates strongly resented the change to a NATO country and also that the meeting included a papal discourse. Recalling the pope's propaganda the previous year, they stayed away from the discourse and the subsequent audience. In his 1952 address to the IAU delegates, the pope expressed himself less openly apologetically than the year before, yet his message was the same: modern astronomy and cosmology indicated the existence of a superior and creative spirit (McLaughlin, 1957, pp. 185-194). The distinguished Soviet astrophysicist Victor Ambartsumian, at the time serving as Vice President of IAU, was among those who strongly disagreed. A convinced Marxist, he subscribed to the view that science and religious faith are irreconcilable.  As he wrote in a paper of 1959 (Graham, 1972, p. 156):

> The history of the development of human knowledge, each step forward in science and technology, each new scientific discovery, irrefutably attests to the truth and fruitfulness of dialectical materialism, … At the same time the achievements of science convincingly demonstrate the complete unsoundness of idealism and agnosticism, and the reactionariness of the religious world view.

Nonetheless, in Rome he assured that astronomers of the world were united in spite of national, religious and political differences. 'We believe that the joint study of such large problems as that of the evolution of celestial bodies will contribute to the cultural rapprochement of different nations, and to a better understanding among them', he said. 'This is our modest contribution to the noble efforts toward maintaining peace throughout the world' (Struve and Zebergs, 1962, p. 32). Ideological tensions eased somewhat after the death of Stalin in 1953, and in 1958 the IAU General Assembly convened in Moscow, largely undisturbed by the Cold War.

## 4. *The Response of Soviet Scientists and Philosophers*

The fear that Spencer Jones aired in his letter to Strömgren, that astronomy and cosmology in the Soviet Union might become 'lysenkoized', was not unfounded but turned out to be exaggerated. The Communist Party was much less interested in cosmology than it was in genetics. There were no purges and no Lysenko in post-World War II Soviet astronomy, as little as there were in physics and chemistry. Ronald Doel and Robert McCutcheon (1995, p. 286) conclude: 'Ideological influences did affect work in the field [of astronomy], as in most disciplines, but the late Stalinist and Krushchev eras were far from tragic periods in the history of Soviet astronomy'.

Yet the Stalinist ideology had a serious effect on cosmology and parts of astrophysics in the Soviet Union, which responded by avoiding ideologically sensitive areas, including cosmology in the style investigated by Western astronomers and physicists. Although there was no ban on this kind of cosmology, political pressure and self-censorship had the consequence that cosmological research was practically non-existent. From 1934 to 1958 there appeared no cosmological models from Soviet scientists corresponding to the kind of models of the universe as a whole discussed in the West (Mikulak, 1958, p. 49; Tropp, Frenkel and Chernin, 1993, p. 225). On the other hand, these models, in most cases based on the equations of general relativity, were well known and sometimes reviewed from a theoretical point of view. The Moscow astrophysicist Abraham Zel'manov wrote on the subject, which was also covered in books by Alexander Fock, an eminent theoretical physicist, and in the influential textbooks on theoretical physics by Lev Landau and Yevgeny Lifshitz. These works dealt with all relativistic models, including those with a beginning in time, but characteristically they were treated from a mathematical point of view and not as possible candidates of the real physical universe.

Following up on the critique by Zhdanov and other party officials, in December 1948 a large number of Soviet astronomers and physicists convened in Leningrad to discuss ideological questions in the astronomical sciences (Prokofieva, 1950; *New York Times*, 14 July 1949). The homogeneous and expanding universe was resolutely criticized as an incorrect extrapolation from observations, and the cosmologists were ordered to find a materialistic interpretation of the redshifts as an alternative to the Western explanation based on the idealistic theory of the expansion of space. The Leningrad science writer V. E. L'lov warned that the relativistic theory of a closed expanding universe was a 'cancerous tumor that corrodes modern astronomical theory and is the main ideological enemy of materialist science'. In the final resolution of the conference, it was stated: 'The reactionary and idealistic "theory" of the expansion of the universe dominates contemporary foreign cosmology. Unfortunately, this anti-scientific theory has penetrated into the pages of our specialized publications … It is indispensable to expose tirelessly this astronomical idealism, which promotes clericalism' (Prokofieva, 1950, p. 19). The themes of the anti-cosmology campaign were not quite new, as ideological critique of relativistic cosmology had been on the agenda also before World War II. In the 1930s the astronomer and party ideologue Vartan Ter-Oganezov, described as something like 'the Lysenko of Soviet astronomy' (Bronshten and McCutcheon, 1995, p. 325), had been particularly active in advocating a dialectical-materialist alternative to the capitalist myth of the expanding universe.

Whereas it was impossible to defend the closed and therefore finite universe, the attitude to cosmic expansion was more mixed. Many of those who discussed the interpretation of the redshifts argued that the phenomenon could be explained on the basis of a static (but infinite) universe, not unlike what Fritz Zwicky and a few other Western astronomers had suggested in the 1930s. Even in the absence of a convincing explanation, they denied that the redshifts proved the cosmic expansion predicted by relativistic cosmology. This expansion is of the entire universe, and the Marxist critics were at most willing to accept the expansion as a local phenomenon. However, other scientists saw no major problem in the relativistic expansion theory, and at the end of the 1950s resistance to it was dwindling. It was now agreed that the expanding universe was not, after all, an 'ideological enemy of materialist science'. This was only the case if it were taken to imply an expansion from a singular state in the past, that is, a finite-age universe of the big bang type.

Stalinism was not only a Marxist-Leninist ideology, it was also xenophobic, anti-cosmopolitan and with a strong element of Russian nationalism. The distinguished astronomer Boris Vorontzoff-Velyaminov attacked Gamow's big bang theory not only because it was unscientific, but also because it was invented by a former Soviet citizen – an 'Americanized apostate' – who had betrayed his socialist fatherland.[5] Scientists in favour of the expanding universe occasionally pointed out that the theory had first been suggested in 1922 by a *Soviet* scientist, Alexander Friedmann, and for this reason alone should not be dismissed as bourgeois idealism. This was what the physicist Dmitri Iwanenko argued at the 1948 Leningrad conference (Prokofieva, 1950, p. 17), but at the time he was an exception and taken to task for speaking favourably of Friedmann's cosmology. The general attitude was to ignore Friedmann, whose embarrassingly idealistic theory was rarely mentioned. For about a decade, he was effectively a 'non-person' (Vucinich, 2001, p. 171).

The universe as described by the steady state theory, since 1948 the main rival to big bang cosmology, was eternal and infinite in size. Moreover, in the West it was widely associated with atheism, an association indirectly supported by Hoyle and a few other steady state theorists. For these reasons one might expect that the Hoyle-Bondi-Gold theory was welcomed in the Soviet Union as agreeing with the requirements of dialectical materialism. But this was not the case at all. While the controversy between the steady state theory and relativistic evolution theories created headlines in England and the United States, it was largely ignored in the Soviet Union. The steady state alternative was known to astronomers and physicists, of course, but it attracted very little attention in scientific and philosophical journals. When it was mentioned, it was to dismiss it as no less reactionary and bourgeois than the big bang theory (Struve and Zebergs, 1962, p. 32; Mikulak, 1958, p. 47). Why?

It appears that there were two main reasons for the unsympathetic response to the steady state theory – apart from its origin in the capitalist world. For one thing, the theory postulated that the universe as a whole was homogeneous in both space and time (the so-called 'perfect cosmological principle'), which was considered even more a priori and idealistic than the ordinary cosmological principle, stating that on a large scale the universe is homogeneous and isotropic. Moreover and probably more seriously, the steady state picture of the universe depended on the hypothesis of continual creation of matter out of nothing, a hypothesis that squarely contradicted the natural philosophy of Engels and Lenin. Whether continually in the form of hydrogen atoms or cataclystically in the form of a big bang, matter creation was seen as an idealistic superstition associated with religion. In 1953 two Soviet astronomers, Boris V. Kukarkin and Alla G. Masevich, explicitly denounced the steady state theory as 'the thoroughly idealistic and absurd theory of the creation of matter' (Graham, 1972, p. 171). They also criticized in similar terms a version of big bang theory proposed by the German physicist Pascual Jordan, one of the founders of quantum mechanics, according to whom entire stars were formed along with the expansion of space. Of course, Jordan's theory was totally unacceptable to defenders of dialectical materialism. Kukarkin and Masevich had attended the IAU congress in Rome and there experienced the pope's apologetic misuse of creation cosmologies. As they saw it, Jordan's theory was eminently suited for this kind of religious exploitation (Kragh, 2004, pp. 183-185).

One way of reconciling an infinitely old universe with the redshifts predicted by relativistic cosmology was to assume a cyclic or oscillating model. Such models were occasionally discussed by Western cosmologists, but they were seriously considered only by a minority, including William Bonnor in England and Herman Zanstra in the Netherlands (Kragh, 2009). On the other hand, among the materialists and socialists in the late nineteenth century ideas of a cyclic universe were very popular and the favoured alternative to the running-down universe associated with theism.

For example, Engels was committed to an eternally cyclic universe. Given the history of this kind of cosmological thinking, it is remarkable that cyclic models were generally dismissed or ignored by Soviet scientists in the 1950s (Graham, 1972, p. 146 and p. 178). They were not considered in agreement with the doctrines of dialectical materialism, for other reasons because a relativistic cyclic universe must be closed. Although time was infinite, space was not, and this contradicted the sanctioned view.

## *5. Metagalaxy or Universe?*

According to the tradition of Western cosmology that can be traced back to Einstein's static and closed world model of 1917, the universe in its *totality* was the proper domain of cosmological research. The solutions of the field equations, supplemented with observational data and uniformity assumptions, described candidates for the entire universe, including regions that cannot be observed even in principle. (Such unobservable regions follow from some expansion models.) This extrapolatory approach to cosmology was sometimes criticized by Western astronomers, who also questioned the cosmological principle of spatial uniformity on which most (but not all) relativistic models relied. In the Soviet Union a similarly critical attitude was not only common, it was on a more fundamental level and until about 1958 it was shared by all astronomers, physicists and philosophers. Moreover, and contrary to the situation in the Western countries, it was often justified in the name of the philosophical system of dialectical materialism.

It was generally agreed that the concept of the universe as a whole was an illegitimate theoretical construct that lacked empirical justification. In some cases, if far from all, it was seen as reflecting the idealism characteristic of capitalist science. Ultimate extrapolations of observations and theories based on the empirically accessible part of the universe were considered unjustified, speculative, and often un-Marxist. Instead of the universe at large, the proper domain of cosmology and cosmogony was held to be the 'metagalaxy', a term typically referring to the assemblage of observable galaxies and clusters of galaxies. Although the concept of a metagalaxy goes back to 1934, when it was introduced by the American astronomer Harlow Shapley, as a substitution for the universe it belongs to the Soviet context in the Stalin and early post-Stalin period.

To the extent Soviet scientists accepted the expansion of the universe – and most did – they thought of the 'universe' as the metagalaxy, which on a cosmological scale is a local object. In a book of 1958, the philosopher Serafim T. Meliukhin wrote about the expansion of the universe as a whole that it was 'antiscientific, contributing to the strengthening of fideism' (Graham, 1972, p. 178). On the other hand, he had no problem with the expansion of the observable part of the universe, the metagalaxy. Among the astronomers, the views of Ambartsumian are of particular interest because of his high status in both national and international science.

Armenian-born Ambartsumian was primarily a theoretical astrophysicist who did very important work in the formation processes of stars, the physics of gaseous clouds and related subjects (Lynden-Bell and Gurzadyan, 1998). As a young man he had worked at the famous Pulkovo Observatory outside Leningrad (St. Petersburg), where in 1935-1936 he became involved in a controversy with the director of the observatory, Boris Gerasimovich. The following year Stalin's Great Terror hit Russian astronomy, resulting in a purge of the Pulkovo astronomers and the execution of several of them; others were imprisoned or just disappeared.

Arrested as an 'enemy of the people', Gerasimovich ended his life before a firing squad. The brilliant Leningrad physicist and cosmologist Matvei Bronstein suffered the same tragic fate. He was falsely charged with being a foreign spy and with 'resolutely opposing materialist dialectics being applied to natural science' (Gorelik and Frenkel, 1994, p. 145). Ambartsumian escaped arrest, but in 1938 he was publicly attacked by L'lov, who called him a 'cleverly masked enemy of Marxism-Leninism' (Eremeeva, 1995, p. 310). As proof of Ambartsumian's crimes, L'lov claimed that he supported Lemâitre's idealistic theory of a created and expanding universe. This was a potentially dangerous accusation, but no action was taken by the political authorities. Quickly changing to a loyal Marxist, in 1940 Ambartsumian became a member of the Communist Party and ten years later deputy to the Supreme Soviet for the Republic of Armenia. Among his numerous later awards and honours were the Lenin Prize, the Stalin Prize and the honorary title of Hero of Soviet Labour.

Ambartsumian was an astrophysicist, not a cosmologist, and his ideas of the universe were influenced both by his background in astrophysics and his adherence to Marxist-Leninist philosophy. He had no problem with either relativistic cosmology or its explanation of the redshifts in terms of a cosmic expansion, but he did object to the extrapolation from the metagalaxy to the universe as a whole (Graham, 1972, pp. 165-171; Vucinich, 2001, pp. 174-176). Likewise, he objected to the extrapolation backwards in time from which some Western cosmologists inferred that the universe had come into existence a finite time ago. Such 'unrestrained extrapolations' were totally unjustified, he argued, for other reasons because they relied on the assumption of a homogeneous universe. According to Ambartsumian, the metagalaxy was far from homogeneous, and there was no reason at all to believe that the universe possessed this property on an even larger scale. His arguments for warning against Western-style cosmology were primarily scientific and methodological, but of course philosophical considerations entered as well. Although not hostile to cosmological models, he believed the relevant data were too few and too uncertain to make them even approximately realistic. 'The character of these models', he said in 1963, 'depends so much on simplifying assumptions that they must be considered far from reality' (Graham, 1972, p. 168). This was not a particularly controversial attitude and neither was it one restricted to Soviet astronomers.

While Ambartsumian's skepticism was not governed by his adherence to the doctrines of dialectical materialism, it was congruent with them. At the 14th International Philosophy Congress held in Vienna 1968, he said: 'The philosophy of dialectical materialism has been assisting and continues to assist many natural scientists, among whom I count myself, in conceptualizing a number of different problems. Of course, this philosophy does not represent a dogma or a universal prescription for all the instances in life' (Ambartsumian, 1969, p. 618). In some cases he supported his views with arguments based on doctrines of dialectical materialism, such as the 'law' of the transformation of quantitative into qualitative changes that Engels had formulated in his *Dialektik der Natur* and which he had inherited from Hegel. Generally Ambartsumian believed that 'the evolution of the universe is cast within the framework of a struggle of dialectically contradictory tendencies', as he phrased it in a publication of 1967 (Vucinich, 2001, p. 175). However, whether dressed in Marxist language or not, his cosmological ideas were not taken very seriously by the majority of Soviet physicists and astronomers. In the West, they were politely ignored.

One of the fundamental problems much discussed by Soviet astronomers and philosophers was the role of philosophy in cosmology, especially with regard to the infinity of the universe. Can such a question ever be answered within the realms of science, or does it need philosophical analysis?[6] Although the infinity of space was never seriously questioned, there was no agreement on the role of philosophy in this and related problems. The discussion in the Soviet Union was to some extent comparable to the one in England, where scientists and philosophers also disagreed about the scientific status of physical cosmology and the necessity of philosophical arguments (see, e.g., Whitrow and Bondi, 1954). The difference was that whereas Western cosmologists referred to philosophy in a general sense, in the Soviet Union the discussion went on within the framework of one particular philosophical school, the one of dialectical materialism.

During the early Cold War period, science under the guidance of Marxism-Leninism was not only defended in the Soviet Union and other socialist countries, but also by some red scientists and intellectuals in the Western world. Cosmology did not create a stir comparable to that of genetics, and only very few Western scientists felt tempted to judge cosmology from an ideological perspective. Among the few was the young French astrophysicist Evry Schatzman, an orthodox Communist with close connections to Moscow.[7] At the request of Kukarkin, whom he had met at the 1952 IAU meeting in Rome, he wrote a lengthy report on cosmogony and cosmology in the Western countries which was published in the Soviet periodical *Voprosy Kosmogonii*. In 1957 the report was turned into a book, *Origine et évolution des mondes*, and in 1966 a revised version of it appeared in an English translation. Apart from occasional references to Engels and ideas based on Marxist philosophy, the book avoided mixing science and politics, and yet it reflected the views that Schatzman shared with his Soviet colleagues. For example, he argued strongly against the heat death, because, 'for an infinite universe, the law of increasing entropy is not valid, either for the universe as a whole or for any infinite part of the universe'. Like Ambartsumian and most other Sovjet astronomers, he dismissed the cosmological principle: 'Very sound physical reasons show that a homogeneous and isotropic universe is a picture which cannot have the slightest connection with reality' (Schatzman, 1966, p. 271 and p. 214). He also resisted the idea of element formation in a big bang, as argued by Gamow and his collaborators, and instead suggested that all the elements, including helium, were produced in stellar nuclear reactions.

## 6. *Revival of Soviet Cosmology*

After Stalin's death in March 1953, science in the Soviet Union changed in several ways, both administratively and as to the relationship between science and ideology, and between basic and applied research. International contacts were revived, so that Soviet membership in international scientific organizations increased from only 2 in 1953 to 42 in 1956, and to 89 in 1960 (Ivanov, 2002, p. 324). On the ideological front, the influence of the party philosophers declined, with most areas of science loosening or abandoning their former links to dialectical-materialist philosophy. The change is clearly visible in the development of astronomy, astrophysics and cosmology in the decade after 1953.

To some extent guided by the official state philosophy, Soviet astronomers in the mid-1950s rejected relativistic evolution cosmologies of the type with a condensed, pre-stellar universe in the past. They also rejected the alternative steady state theory, and more generally they questioned the very idea of a scientific model for the entire universe. Thus, their contributions were essentially negative and critical, while there were no attempts to formulate an independent cosmology in agreement with the guidelines of dialectical materialism. Soviet studies included the metagalaxy and 'cosmogonies' of local objects such as the solar system and galaxies, but these were not comparable to cosmology as cultivated in the West. By and large, Sovjet astronomers conformed to the dogmas of the Communist Party by giving up the study of the universe as a whole.

A meeting of the Commission for Cosmogony of the USSR Astronomical Council in late 1956 gives an impression of the weaknesses in Soviet cosmology but also of the emerging recognition that a break with the unfruitful attitude of the past was needed. According to the report prepared by Masevich (1957), the absence of translations of foreign monographs and research papers in cosmology was a problem. And, without denouncing in any way the value of dialectical materialism: 'It is important not to introduce simplifications and dogmatism'. Ambartsumian and Zel'manov both admitted that 'cosmological problems are somewhat neglected in the USSR while a considerable number of papers are appearing abroad'. While the first three volumes of *Astronomicheskii Zhurnal* (1957-1959) included no papers under the category 'Cosmology and Cosmology', during the 1960s the number increased to an average of seven papers per volume. A change was under way, not only quantitatively but also qualitatively and with respect to Soviet cosmology's relation to political ideology. The change can be followed by comparing three decennial jubilee volumes on astronomy written in 1947, 1957, and 1967, respectively (Tropp, Frenkel and Chernin, 1993, pp. 225-226). Likewise, a comparison of the entries on cosmology and cosmogony in the second and third editions of the *Great Soviet Encyclopedia*, published 1950-1958 and 1969-1978, respectively, demonstrates the dramatic change.

'In accordance with the presently observed expansion of the universe', reads a papers of 1962, 'it is deemed probable that in the earlier stages of the evolution of the universe there existed a homogeneous isotropic Friedmann nonstationary solution with the density of matter decreasing from an infinite value at the initial instant' (Zel'dovich, 1963, p. 1102). In regard of the traditional hostility to homogeneous finite-age models, these were remarkable words from a prominent Soviet scientist; and they were no less remarkable in the light of the very limited interest that Western scientists at the time paid to models of the big bang type. The words came from Yakov Zel'dovich, a rising star in Soviet astrophysics and cosmology who had originally specialized in nuclear physics and been a leading member, together with Igor Kurchatov and Andrei Sakharov, of the Soviet nuclear bomb programmes. Three times a Hero of Socialist Labour and the recipient of a Lenin Prize and four Stalin Prizes, he was no less a heavyweighter in the Soviet science system than Ambartsumian.

The kind of cosmology that Zel'dovich and his pupils developed in the 1960s was squarely within the framework of Western mainstream cosmology. He was convinced that cosmology was a theory of the universe as a whole and that it had to be based on Einstein's equations. Like his colleagues in the West, he came to the conclusion that for some ten billion years ago the density of the universe had been infinite or nearly so, and he refrained from speculating about the ultimate creation, had there ever been one. Uninterested in the doctrines of dialectical materialism, his important publications from the 1960s were purely technical and without references to the philosophical discussions that a decade earlier had been common in parts of Soviet cosmology. *Relativistic Astrophysics*, a book by Zel'dovich and his principal collaborator Igor Novikov written in Russian between 1965 and 1967, was one of the very first monographs to give a comprehensive and modern account of all aspects of modern cosmology. Very little in it revealed that it was a product of the Soviet Union and not, for example, the United States.

## *7. A Note on Red China*

While Soviet science was gradually depoliticized during the 1950s and 1960s, the *de facto* ban on cosmology in the Western sense went unchallenged in the People's Republic of China, where radical Maoist ideologues developed their own version of dialectical materialism.[8] The ideological interference with cosmological theory took a new turn during the Cultural Revolution in Mao Zedong's empire, when relativistic cosmology for a period was declared a reactionary and anti-socialist pseudo-science. Fang Lizhi, a physicist who had changed his research interest from solid-state physics to astrophysics, got caught up in the frenzy of the Cultural Revolution (Williams, 1999). He was arrested and imprisoned as a class enemy and 'rightist', but was able to resume his scientific career. In 1972 he published a theoretical paper in a new physics journal on big bang cosmology and the cosmic microwave radiation, the first of its kind in the People's Republic. Enraged radical Marxists immediately rallied against Fang's heresy and its betrayal of the true spirit of proletarian science. According to one critic, Li Ke, the big bang theory was nothing but 'political opium' (Cheng, 2006, p. 135).

During the next couple of years, some thirty papers were published against the bourgeois big bang theory and cosmology in general. As late as 1976, the journal *Acta Physica Sinica* carried an article that warned against 'the schools of physics promoting a finite universe [which] are linked up with all sorts of idealist philosophy, including theology'.[9] The article summarized what was wrong with this kind of theory:

> Materialism asserts that the universe is infinite, while idealism advocates finitude. At every stage in the history of physics, these two philosophical lines have engaged in fierce struggle … with every new advance in science the idealists distort and take advantage of the latest results to 'prove' with varying sleights of hand that the universe is finite, serving the reactionary rule of the moribund exploiting classes … We must ferret out and combat every kind of reactionary philosophical viewpoint in the domain of scientific research, using Marxism to establish our position in the natural sciences.

Similar denunciations of modern cosmology as antagonistic to Maoist thought appeared in many other journals and in the national news media until the end of 1976. The party line was to deny cosmology scientific legitimacy, much like materialists had argued in the nineteenth century and as Soviet ideologues had argued in the Stalin era. Questions of the universe at large could not be answered scientifically, but only on the basis of 'the profound philosophical synthesis' of Marxism-Leninism: 'The dialectical-materialist conception of the universe tells us that the natural world is infinite, and it exists indefinitely. The world is infinite. Both space and time are boundless and infinite' (Fang, 1991, p. 311). The campaign against Fang and big bang cosmology was closely connected to the anti-Einstein campaign that started in 1968 and culminated in the early 1970s (Hu, 2005).

The Chinese anti-cosmology campaign came at a time when the Cultural Revolution was on decline, and in 1975 Fang and his colleagues were allowed to defend themselves. 'Whether the big bang is a correct theory or not', they proudly stated, 'recent developments such as radiotelescopy had made cosmology an experimental science, to be approached though the usual scientific methods rather than through philosophical discourse' (Hu, 2005, p. 168). Still, they were careful to point out that the scientific methods were in agreement with Chairman Mao's dialectical views on nature. At a later occasion, after the fall of the Gang of Four in 1976, Fang expressed himself in a less conciliatory tone. He recalled about the earlier battle (Williams, 1990, pp. 466-467):

> The so-called 'Big Criticism of Science' became the highest arbiter of scientific right or wrong. No reliance was placed on experiment, and all scientific controversies were treated according to certain a priori principles. Big-bang cosmology, and alas the whole of modern cosmology, received the theoretical equivalent of a death sentence at its hands.

Although the rule of ideology over modern cosmology came to an end in late 1976, the core of Maoist cosmology remained intact: the spatial and temporal infinity of the universe, and the guiding role of Marxist-Leninist-Maoist philosophy in science, continued to govern much of scientific thinking in Red China. As to Fang Lizhi, his scientific unorthodoxy led him to political unorthodoxy. He developed into China's most prominent political dissident, and after the Tianmen Square massacre of June 1989 he escaped to the United States (Fang, 1991).

## *8. Conclusion*

Contrary to sciences such as chemistry, medicine, geology and physics, cosmology has no technological or military applications whatever. In Marxist terminology, it is not a productive force. All the same, because of its traditional association to philosophical and religious world views it has often played a political role, however indirect. As a science of the universe at large, cosmology experienced a minor revolution in the late 1940s, just at the time when the Cold War intensified and threatened to develop into a hot war. While the new cosmological theories, the big bang theory and the rival steady state theory, attracted little political attention in the West – although they did attract some religious attention – in Stalin's Soviet Union cosmology came to be seen as an ideological battleground of great importance. There were many discussions of cosmological subjects among astronomers and philosophers, but for more than a decade they were constrained by the requirements of dialectical-materialist thinking. By ideological decree, the universe could not be finite in either space or time, and it could not evolve irreversibly towards an equilibrium state. More generally, the universe as a whole could not be the subject of science, but only of philosophy in the form of Marxism-Leninism.

The case discussed in this paper, and the corresponding case of Maoist doctrines of cosmology in Red China, is unique in the post-World War II history of the physical sciences. Although specifically related to the political context in the Communist countries, there are some similarities to how cosmology was discussed by Western scientists and philosophers. Some of the issues of contention, such as the legitimacy of matter creation and the scientific status of the universe as a whole, were the same, and yet they were discussed in an atmosphere essentially free of fixed philosophical doctrines. I think Graham (1972, p. 194) exaggerates the similarity when he suggests that the Soviet cosmologists' efforts to fit cosmology into the system of dialectical materialism 'were not so dissimilar, in essence, from the efforts of many non-Soviet philosophers or scientists'. As it came to be admitted by Soviet scientists, the preconception that the science of the cosmos must conform to the dogmas of Marxist thought was a mistake that retarded the development of astronomy and cosmology in the Communist countries. On the other hand, the damage caused by the excessive politicization was temporary only, as witnessed by the remarkable progress beginning in the 1960s. The case here examined not only illustrates the harmful effects of imposing ideological views on science, it also exemplifies what were after all the strengths of the Soviet science system, or what Kojevnikov (2004) calls 'Stalin's Great Science'.

NOTES

[1] Yet another case of relevance to the one of cosmology is the Russian opposition to the continental drift theory and its later development into plate tectonics (Wood, 1985, pp. 210-223; Khain and Ryabikhin, 2002). At least in part, the opposition was ideologically motivated.

[2] *Voprosy Filosofii*, no. 1 (1947), p. 271, as translated in Tropp, Frenkel and Chernin (1993, pp. 223-224). For a full translation of Zhdanov's address, see Wetter (1953, pp. 594-616). Although the universe proposed by the English astrophysicist E. Arthur Milne was of finite age, it was spatially infinite. While Zhdanov and other Soviet philosophers labelled him a reactionary idealist, according to the Marxist biologist John B. S. Haldane his cosmology was 'beautifully dialectical' and in harmony with Marxist thought (Kragh, 2004, p. 221). Contrary to what Zhdanov asserted, the cosmological models investigated in the capitalist countries did not generally assume the universe to be limited in space and time.

[3] Plotkin (1951). Discussions of this kind, sometimes influenced by philosophical or emotional desires, were not unknown in the Western countries. For example, Milne argued that the notion of entropy increase was inapplicable to the universe at large and that a heat death would probably never occur (Kragh, 2004, pp. 202-204). A devout Christian, Milne thus arrived at the same conclusion as most atheists and socialists, an infinite universe with an unlimited future.

[4] For example, William Bonnor (1964, p. 117), an English atheist cosmologist critical to the big bang theory: 'The underlying motive is, of course, to bring in God as creator.' Somewhat similar statements can be found in Hoyle, the philosopher Stephen Toulmin, and several other Western commentators (see examples in Kragh, 2004).

[5] According to B. Vorontzoff-Velyaminov, *Gaseous nebulae and new stars* (in Russian), Moscow: USSR Academy of Sciences, 1948, as quoted in Otto Struve's review essay in *Astrophysical Journal*, 110 (1949), 315-318.

[6] It may not be impertinent to mention that this is still an unsolved problem. Some modern cosmologists argue that the infinity of space is a philosophical and not a scientific question.

[7] Schatzman (1996) gives a fascinating insight in the intellectual attraction of Marxism in the early Cold War era. Schatzman describes his discovery of Marxism-Leninism in the 1940s 'as if I had taken holy orders' and 'it was like sunshine enlightening my life' (p. 14). Only in 1956 did he realize the reality of the repressive Soviet system, and in 1959, while still subscribing to the dogmas of Marx and Engels, he left the Communist Party. It took him another decade to realize that 'knowledge of society is not the same as knowledge of science' (p. 18). On the Marxist milieu in French science to which Schatzman belonged, see Cross (1991, pp. 747-750).
[8] My brief account is only to make aware of the Chinese case and its obvious similarity to the one of the Soviet Union in the earlier phase of the Cold War. Much more detailed analyses can be found in Williams (1990), Williams (1999) and Cheng (2006).
[9] The article, written by Liu Bowen, is translated in Fang (1991, pp. 309-313), from where the quotations are taken.